\documentclass[preprint2]{aastex6}
\usepackage{epsfig}
\usepackage{amsfonts,amssymb,amsmath}

\usepackage{caption}
\usepackage{subcaption}

\begin{document}

\title{Brightest Cluster Galaxies and Intra-Cluster Light: Their Mass Distribution in the Innermost Regions of Groups and Clusters}
\author{E. Contini$^{1,2}$ and Q. Gu$^{1,2}$}

\affil{$^1$School of Astronomy and Space Science, Nanjing University, Nanjing 210093, China; {\color{blue} emanuele.contini82@gmail.com, qsgu@nju.edu.cn}}
\affil{$^2$Key Laboratory of Modern Astronomy and Astrophysics (Nanjing University), Ministry of Education, China}

\begin{abstract} 
We improve the model presented in \cite{contini20} that describes the radial mass distribution of brightest cluster galaxies (BCGs) and the diffuse component also known 
as intra-cluster light (ICL), by assuming that the global BCG+ICL radial mass distribution follows the sum of three profiles: a Jaffe and an exponential profiles for the bulge 
and disk of the BCG, respectively, and a modified version of an NFW profile for the ICL. We take advantage of a wide sample of BCG+ICL systems simulated with our state-of-art 
semi-analytic model to: (a) investigate the reliability of our BCG+ICL distribution by looking at several scaling relations between the BCG+ICL stellar mass within different 
apertures and the total BCG+ICL/halo mass, at different redshift; (b) make a prediction of the distance where the radial distribution transitions 
from BCG to ICL dominated. We find that our model nicely reproduces all the observed scaling relations investigated at the present time with a compelling degree of precision, but 
slightly biased-low with respect to observations at higher redshifts ($z\gtrsim 0.5$). The transition radius predicted by our model is in good agreement with recent observational 
results, and spans a range between $\sim 15$ kpc and $\sim 100$ kpc. It mostly depends on the morphology of the BCG, whether it is bulge or disk dominated, on the amount of ICL 
with respect to the bulge and/or disk, and on the dynamical state of the group/cluster.
\end{abstract}


\section[]{Introduction} 
\label{sec:intro}

Discovered for the first time by \cite{zwicky37}, the Intra-Cluster Light (hereafter ICL) is a diffuse light in groups and clusters from stars that are not gravitationally bound to any galaxy, and which can 
extend to several hundreds kpc (\citealt{gonzalez05,zibetti05,krick07,gonzalez13,giallongo14,zhang19,henden20} and references therein). In the recent past, many attempts, both theoretical and observational (\citealt{puchwein10,rudick11,presotto14,contini14,burke15,demaio15,edwards16,tang18,montes19,kluge20,contini20,spavone20,furnell21}, 
and references therein), have been made in order to study the formation and evolution of this peculiar component, and to highlight its importance in the context of the growth of galaxy clusters and their dynamical state. 

Most of the above quoted studies (and many others) agree on the fact that the ICL forms mainly via tidal stripping of intermediate/massive galaxies and mergers of galaxies orbiting around the potential well of 
the cluster (see \citealt{contini18,contini19} and references therein for a detailed discussion, or \citealt{mihos17,montes19b} for reviews). From the theoretical point of view, it is almost trivial\footnote{For 
the sake of truth it is not very straightforward to distinguish between merger and stripping channels in the very moment two galaxies are starting to merge simply because there is not a clear definition of when a merger 
starts. This isssue is particularly important in simulations.} to isolate the contribution of each channel of the formation of the ICL. Observationally speaking, it is possible to look at properties of the ICL such as 
metallicity and color and compare them with the typical metallicity or color of galaxies (e.g., \citealt{demaio15,morishita17,iodice17,montes18,iodice20}) to understand what channel contributes the most.

Regardless of the processes that bring to the formation of the ICL, or the main contributors to it, in the very last years a few authors (e.g., \citealt{montes19,kluge20,alonso20,contini20,deason21,poliakov21}) have 
focused on the link between the ICL and the dark matter distributions, to figure out whether or not the ICL can be used as a tracer of the dark matter in clusters. By using a sample of six clusters from the Hubble Frontier Fields 
(\citealt{lotz17}), \cite{montes19} compared the bi-dimensional distribution of the dark matter with that of the ICL making use of the Modified Hausdorff distance (MHD), which gives the idea of how far two distributions 
are from each other. They found that the average distance between the ICL and the dark matter within 140 kpc from the center is around 25 kpc, which is a very short distance considering the size of typical galaxy 
clusters. \cite{kluge20} investigated on the ICL-cluster alignment with a sample of around 50 local clusters and focused on the ICL-cluster and brightest cluster galaxy (hereafter BCG)-cluster alignments. They found 
that the ICL is better aligned than the BCG with the host cluster in terms of both position and centering. These results strongly suggest that the ICL follows the global dark matter distribution and can be used as a 
tracer of it. Similarly, \cite{alonso20} used the Cluster-EAGLE simulations (\citealt{barnes17,bahe17}) to test the results mentioned above. By using the same procedure used in Montes et al., they found that the stellar 
mass distribution follows that of the total, including dark matter, but their radial profiles differ substantially. 

In our last paper (\citealt{contini20}, hereafter CG20), we developed an ICL profile to study its distribution in galaxy groups and clusters for a sample of more than 300 BCGs, whose properties 
were obtained by means of our semi-analytic model. We started from the idea, supported by the recent observational results mentioned above, that the ICL can follow the dark matter distribution. Hence, we assumed that the 
ICL follows an NFW (\citealt{navarro97}) profile and added a new parameter, that we called $\gamma$, which is linked to the halo concentration by the relation $c_{ICL}=\gamma c_{DM}$. The parameter $\gamma$ has been then 
set to reproduce the observed $M^*_{100}-M_{500}$ relation, between the BCG-ICL stellar mass in the innermost 100 kpc and the halo mass. By testing our model against several observational and theoretical results, we 
found that $\gamma=3$ is required to match the observed $M^*_{100}-M_{500}$ at $z=0$, while larger values of $\gamma$ are needed at higher redshift, with also hints for a possible halo mass dependence. We then 
concluded that an NFW like profile is a good description of the ICL distribution in groups and clusters, making the ICL a reliable tracer of the dark matter.

The idea of adopting an NFW profile to describe the distribution of the ICL is not original. For example, \cite{zibetti05} found that the observed ICL profile can be approximated by an NFW. They analysed the spatial 
distribution and color of the ICL taking advantage of a sample counting almost 700 galaxy clusters between $z\sim 0.2$ and $z\sim 0.3$ selected from the first release of the Sloan Digital Sky Survey (SDSS-DR1, 
\citealt{abazajian03}), finding that the radial distribution of the ICL is more centrally concentrated than that of the galaxies, a result which well agrees with our assumption. Recently, some evindence come from 
the study of \cite{zhang19}. These authors use around 300 galaxy clusters between $z\sim 0.2$ and $z\sim 0.3$ selected from the Dark Energy Survey (DES, \citealt{mcclintock19}) and focus on the detection of the ICL. 
Among various interesting results, they find that the ICL radial profile is self-similar and scales with the virial radius of the clusters and that the ICL brightness is a good tracer of the cluster radial mass 
distribution, in agreement with the studies quoted above. There is evidence also from the theoretical side. Just to mention one, \cite{pillepich18} studied the properties of the ICL as predicted by the IllustrisTNG 
hydrodynamical simulation and found that the radial distribution of the ICL is as shallow as that of the dark matter, meaning that the ICL can effectively trace the overall cluster mass distribution.

The aim of this paper is to further improve the model developed in CG20, where BCGs and ICL have been treated as separated components when we focused on the $M^*_{100}-M_{500}$ relation, but BCGs have been assumed to be 
entirely contained within 100 kpc, a fair assumption in many cases. Here, we treat a BCG as a two-component system, bulge and disk, and model each of them in order to get their mass within any given distance from the 
center. This approach allows us to enter the very innermost regions of the clusters where BCGs are placed, and compare our model predictions with a variety of observed relations. The ICL is still assumed to follow a
modified NFW profile as in CG20, and we focus mainly at the present time given the fact that, in CG20, we found the best match between our predictions and observed relation at $z=0$. Once we show that our improved
model matches a plethora of observed results, we use it to investigate the differential mass distribution as a function of clustercentric distance of the three component, bulge, disk and ICL, for different case studies. 
We will show that, despite common assumptions frequently used, the ICL can be the dominant component even at very low distances from the cluster centre.

The paper is organized as follows. In Section \ref{sec:methods} we briefly describe our semi-analytic model and the profile used to describe any single component. In Section \ref{sec:results}  we present our analysis, 
which will be fully discussed in Section \ref{sec:discussion}. In Section \ref{sec:conclusions} we sum up our main conclusions. Throughout this paper we use a standard cosmology, namely: $\Omega_{\lambda}=0.76$, 
$\Omega_{m}=0.24$, $\Omega_{b}=0.04$, $h=0.72$ , $n_s =0.96$ and $\sigma_{8}=0.8$. Stellar masses are computed with the assumption of a \cite{chabrier03} Initial Mass Function (IMF), and all units are $h$ 
corrected.

\section{Methods}  
\label{sec:methods}

We make use of the semi-analytic model developed in \cite{contini14}, and further improved in \cite{contini18,contini19}, which contains a full description of the formation and evolution of the ICL. For a detailed description 
of the mechanisms that bring to the formation of the ICL, we refer the reader to the papers quoted above, while below we give a short summary of the main features of the semi-analytic model (hereafter SAM) in the context
of this work.

As most of nowadays SAMs, ours runs on the merger trees of N-body simulations. Mergers trees contain all the information about the formation and growth of dark matter haloes and subhaloes and constitute the input 
of our SAM, which, according to the initial conditions (cosmological parameters) of the numerical simulations, places galaxies in haloes and follow them along time by considering the most important physical mechanisms 
in galaxy formation, i.e. gas cooling, star formation, SN and AGN feedback, mergers and stellar stripping.

In this work we use the merger trees extracted from the same set of high-resolution simulations used in CG20, whose details can be found in \cite{contini12}. Our sample accounts for 361 groups and clusters at $z=0$ with 
a virial mass ($M_{200}$) that spans over two orders of magnitude, from $10^{13} \, M_{\odot}/h$ and up to more than $10^{15} \, M_{\odot}/h$.

As in CG20, we model the formation of the ICL by making use of the model "Tidal Radius" described in \cite{contini14} and further improved in \cite{contini18}. The prescription is applied to satellite galaxies orbiting around 
central galaxies. The SAM derives a tidal radius $R_t$ for each satellite galaxy, which is assumed to be a two-component system, bulge and disk. If $R_t$ is contained within the bulge, the satellite is assumed to be 
completely disrupted, while if $R_t$ is larger than the bulge but smaller than the size of the disk, the mass in the shell between $R_t$ and the radius of the disk is stripped. In both cases, all the stripped stellar mass is added 
to the ICL component associated with the BCG. Another prescription, which considers the stellar mass of satellites stripped during mergers, is implemented and accounts for the so-called merger channel. At each merger,
a percentage of 20\% of the stellar mass of the satellite is added to the ICL component that is associated with the BCG, and the rest of the mass ends up to the BCG stellar mass. 

The details of the other minor assumptions concerning the description of the mechanisms that brings to the formation of the ICL can be found in the papers quoted above. Below, we describe how we model the mass 
distribution of BCGs, bulge and disk, and the mass distribution of the ICL.

\subsection{Bulge and Disk Profiles}
\label{sec:bcg_profile}

We assume a BCG to be a two-component system: bulge and disk. The bulge is assumed to follow a Jaffe profile (\citealt{jaffe83}):
\begin{equation}\label{eqn:jaffe1}
\rho_J (r) = \frac{M_J r_j}{4\pi r^2 (r_j +r)^2}  ,
\end{equation}
where $M_J$ is the total mass of the bulge and $r_J$ the scale length. By integrating, the mass contained within a sphere of radius $r$ is given by
\begin{equation}\label{eqn:jaffe1}
M_J (r) = \frac{M_J r_j}{r_j +r}  .
\end{equation}

For the disk, we assume an exponential profile and its mass within a given radius $r$ is given by:
\begin{equation}\label{eqn:disk}
M_D (r) = M_{D, tot} \left[1-\left(1+\frac{r}{R_D} \right)e^{-r/R_D} \right] ,
\end{equation}
where $M_{D, tot}$ is the total stellar mass in the disk, and $R_D$ the scale radius of the disk. By splitting the BCG in the two components, we are able to drop the assumption made in CG20, that is the stellar mass
of the BCG is entirely contained in the innermost 100 kpc. With this approach, we can probe distances much below 100 kpc.

\subsection{ICL Profile}
\label{sec:icl_profile}

We model the mass profile of the ICL exactly as in CG20, by assuming a modified version of the NFW profile. The reason behind our choice is fully discussed in CG20 and in Section \ref{sec:intro} of this paper.
The NFW profile reads as follows:
\begin{equation}\label{eqn:nfw}
\rho (r) = \frac{\rho_0}{\frac{r}{R_s}\left(1+\frac{r}{R_s}\right)^2} \, ,
\end{equation}
where $\rho_0$ is the characteristic density of the halo at the time of its collapse and $R_s$ is its scale radius. The concentration of a dark matter halo is defined as
\begin{equation}\label{eqn:concentration}
c = \frac{R_{200}}{R_s} \, ,
\end{equation}
where $R_{200}$ is the virial radius of the halo.
The modification we made in CG20 to the classic NFW profile described in Equation \ref{eqn:nfw} is by introducing a free parameter in Equation \ref{eqn:concentration}. We define the ICL concentration as
\begin{equation}\label{eqn:gamma}
c_{ICL} = \gamma \frac{R_{200}}{R_s} \, .
\end{equation}

In CG20 we argued that the parameter $\gamma$ could be a function of halo properties such as the halo mass, and redshift. We set $\gamma$ by matching the observed $M^*_{100}-M_{500}$ at different redshifts 
(not considering any halo dependence) and we found that $\gamma =3$ at $z=0$ and higher values at higher redshifts, which means that the ICL is more concentrated than the dark matter halo in which it is contained.
As mentioned in CG20, this is in good agreement with recent theoretical works (e.g. \citealt{contini18,pillepich18}) and observational studies (e.g. \citealt{montes19}).

\section{Results}
\label{sec:results}

In this section we test our model of BCG-ICL in the innermost regions of clusters, below 100 kpc, by looking at the relation between $M_X$ (where $X$ is a given distance from the centre) and the halo/BCG mass, and
their fraction with respect to the total (Section \ref{sec:model_set}). We show that our model compares well with observational results, and we use the success of our model in reproducing those observed quantities to 
investigate the mass distribution of each component as a function of distance from the cluster centre for a few case studies (Section \ref{sec:mass_distr}). 

\begin{figure}[hbt!]
\includegraphics[scale=0.90]{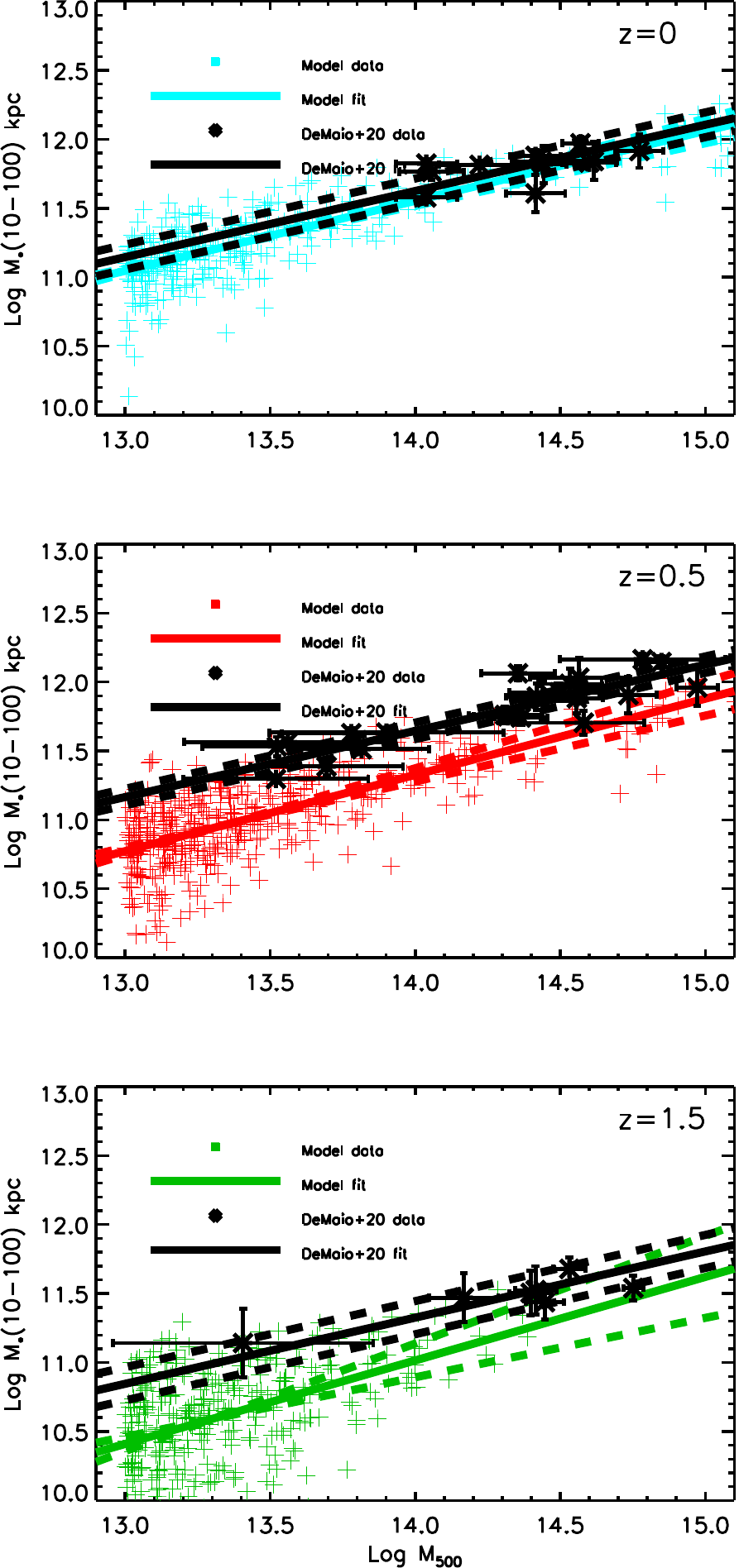}
\centering
\caption{Relation between the BCG-ICL stellar mass enclosed between 10-100 kpc and the halo mass $M_{500}$, for groups and clusters at different redshifts as indicated in the legend. Dashed lines represent $\pm 3 \sigma$ 
scatter around the average relations, marked with solid lines. The model predictions are compared with the observed set of data by \cite{demaio20}, with $\gamma =3$ at $z=0$ and $\gamma=5$ at higher redshifts. Our model agrees well with the observed 
data in the local Universe, and fairly well at higher redshifts, although our predictions are somewhat biased low with respect to observations.}
\label{fig:m10_100_halo}
\end{figure}

\subsection{Model-Observation Comparisons}
\label{sec:model_set}

In Figure \ref{fig:m10_100_halo} we show the predicted $M_{10-100}-M_{500}$ relation, i.e. the BCG-ICL stellar mass between 10-100 kpc and the halo mass at different redshifts (different panels),
compared with the observed relation by \cite{demaio20} at similar redshift. As explained above, the values of $\gamma$ used are $\gamma =3$ at $z=0$ and $\gamma =5$ at higher redshifts. This figure 
is similar to Figure 2 in CG20, with the exception that the first 10 kpc are not considered. As in CG20, we find a reasonable agreement with the observed relations, in particular at the present time.
Nevertheless, it appears clear that our model tends to underpredict, although for $\sim 0.3$ dex, the observed relation at $z \geq 0.5$. The degree of the agreement at $z=0$, instead, is definitely 
excellent as it is in Figure 2 of CG20 where we account for also the first 10 kpc.

\begin{figure}[hbt!]
\includegraphics[scale=0.75]{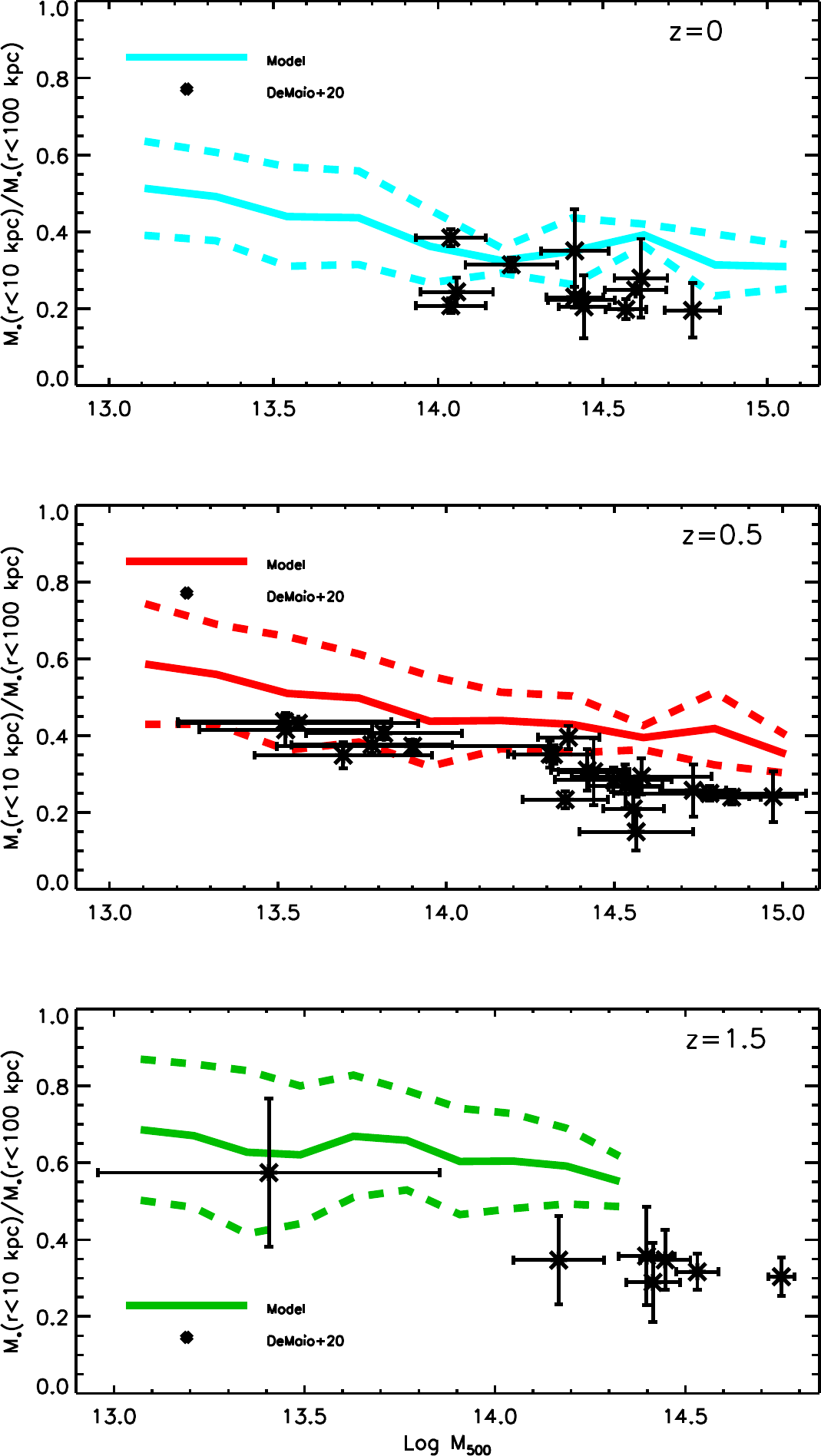}
\centering
\caption{Ratio between the BCG-ICL stellar mass enclosed within 10 kpc and that within 100 kpc as a function of the halo mass $M_{500}$, for groups and clusters at different redshifts as 
indicated in the legend. Dashed lines represent $\pm 1 \sigma$ scatter around the average relations, marked with solid lines. As in Fig \ref{fig:m10_100_halo}, the model predictions are compared with 
the observed set of data by \cite{demaio20}, with $\gamma =3$ at $z=0$ and $\gamma=5$ at higher redshifts. Similarly to Fig \ref{fig:m10_100_halo}, our model predictions compare fairly well with observations, 
particularly at the present time. Regardless of the redshift, a weak trend is visible: less massive haloes contain more BCG-ICL mass in the innermost 10 kpc than more massive ones.}
\label{fig:ratio_halo}
\end{figure}

In order to investigate whether our model overpredicts the amount of stellar mass in the first 10 kpc, in Figure \ref{fig:ratio_halo} we plot the ratio between the stellar mass within 10 kpc and that 
within 100 kpc, as a function of the halo mass and at different redshifts (different panels). Once again, our predictions (color lines) are compared with the observed data by De Maio et al. (black symbols).
The agreement looks better than that found in Figure \ref{fig:m10_100_halo}, at least at $z\leq 0.5$, although it is clear that our predictions are somewhat higher than the observed data. Considering the 
fact that the predicted mass within 100 kpc is either similar or lower than observed, it suggests that our model predicts a slightly higher amount of stellar mass in the first 10 kpc. Moreover, regardless 
of the redshift, less massive haloes have higher $M_{10}$ than more massive ones. This results from the fact that the parameter $\gamma$ is fixed with halo mass, but less massive haloes are intrinsecally 
more concentrated than more massive ones. Hence, on average, the percentage of ICL within the first 10 kpc is higher for less massive haloes.

\begin{figure}[hbt!]
\includegraphics[scale=0.75]{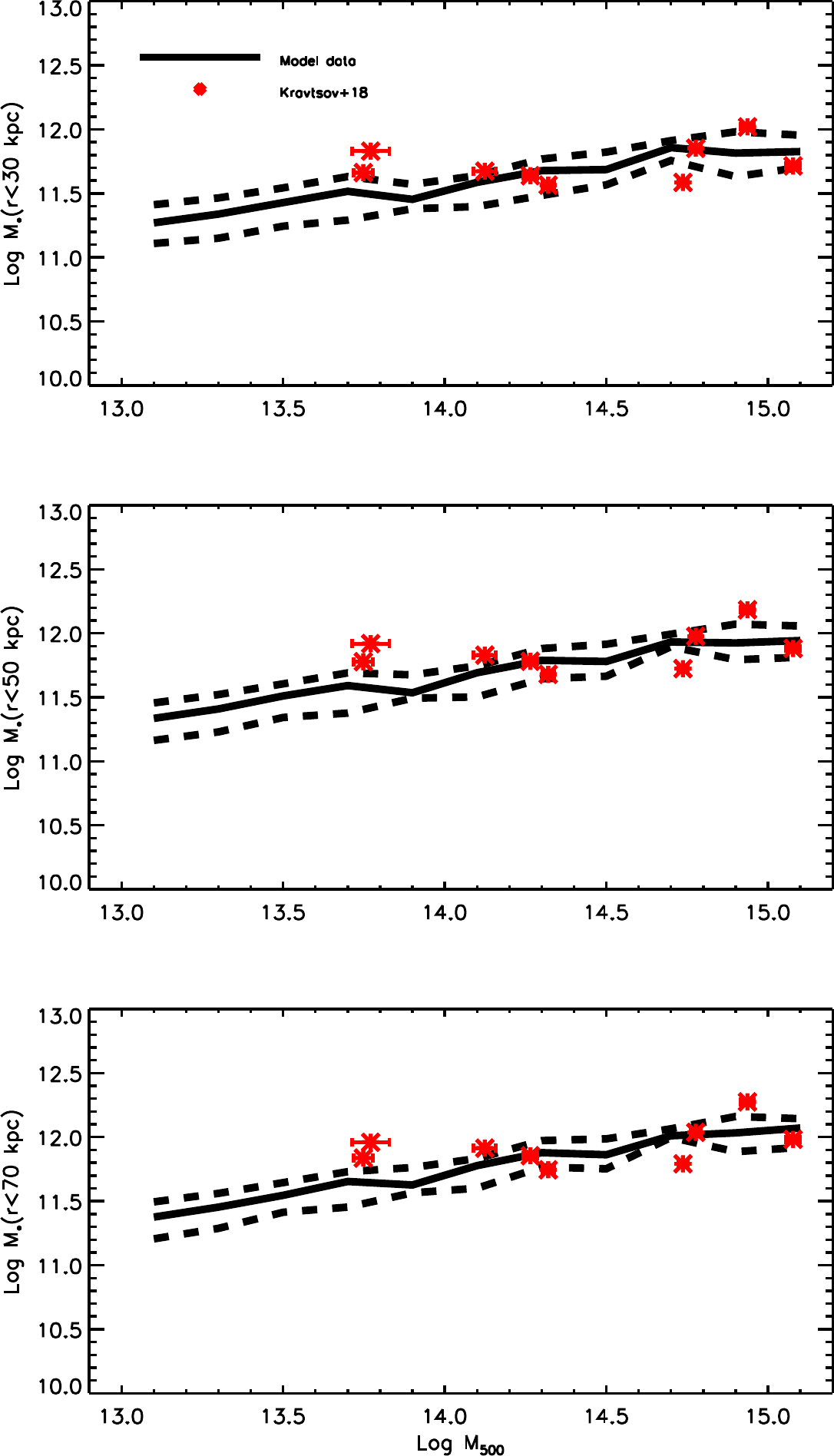}
\centering
\caption{BCG-ICL stellar mass within different apertures (30, 50 and 70 kpc) as a function of the halo mass $M_{500}$, in the local universe. Dashed lines represent the 16-th and 84-th percentiles, while  
solid lines represent the median of the distributions. Our model predictions (black lines) are compared with observed data (red stars) by \cite{kravtsov18}. Our model is able to predict the right amount of BCG-ICL mass 
within different apertures in a wide range of halo mass.}
\label{fig:mXX__halo}
\end{figure}

\begin{figure}[hbt!]
\includegraphics[scale=0.75]{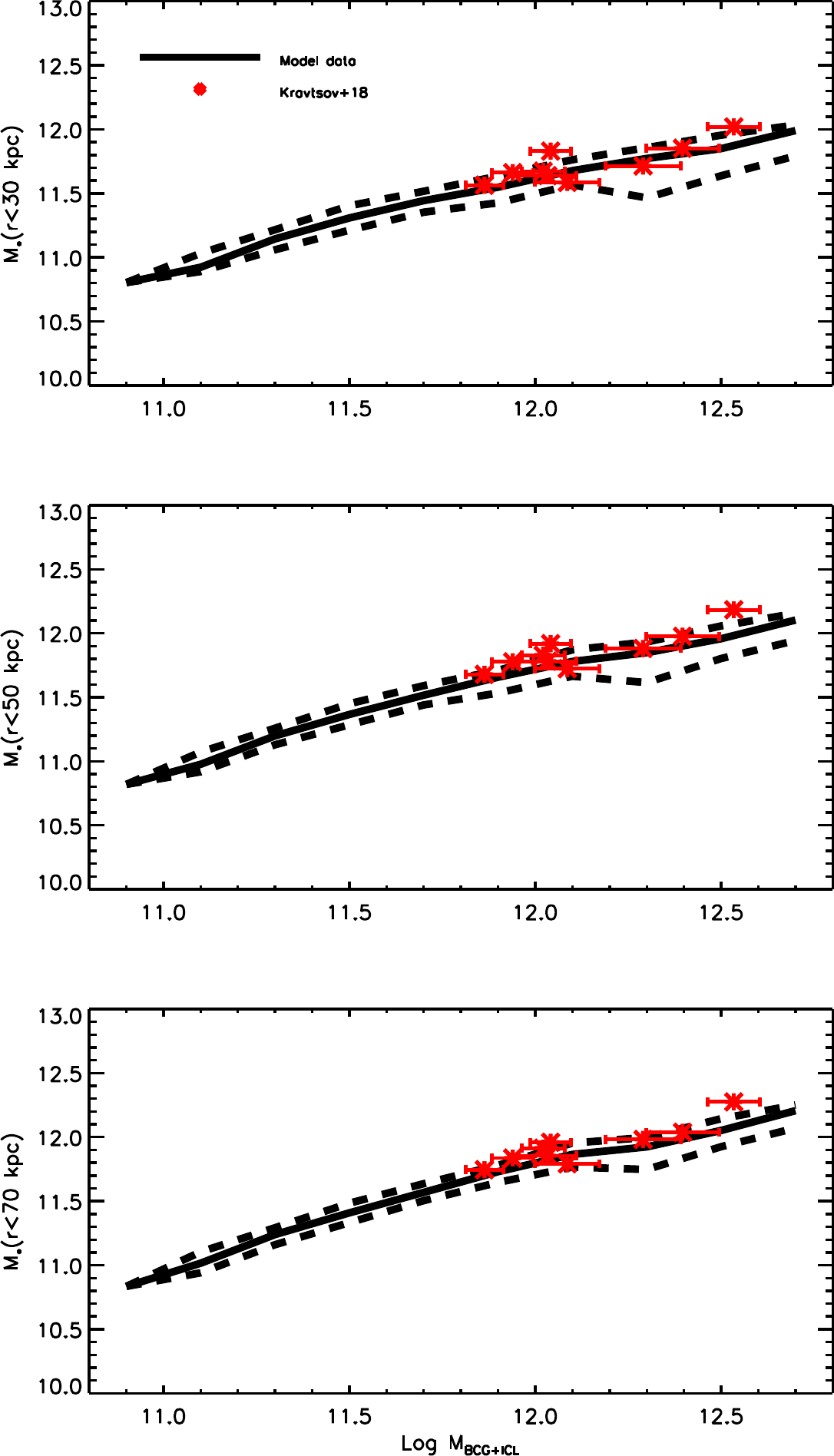}
\centering
\caption{BCG-ICL stellar mass within different apertures (30, 50 and 70 kpc) as a function of the amount of BCG-ICL mass within a given radius (see text for 
further details), at $z=0$. Dashed lines represent the 16-th and 84-th percentiles, while solid lines represent the median of the distributions. Our model predictions (black lines) are in good 
agreement with the observed data (red stars) by \cite{kravtsov18}.}
\label{fig:mXX_rout_miclbcg}
\end{figure}

To give more credit to our model, we now investigate on the BCG-ICL mass within different apertures, and on their ratio over the total, as a function of halo and BCG mass. Since observations are so far better 
reproduced by our model at $z=0$, we will focus only at the present time. In Figure \ref{fig:mXX__halo} we plot the BCG-ICL mass within 30, 50 and 70 kpc as a function of halo mass, as predicted by the model 
(black lines), and as observed (red symbols) by \cite{kravtsov18}. Our model can predict the right amount of BCG-ICL mass within different apertures for a wide range of halo mass. 
Similarly, Figure \ref{fig:mXX_rout_miclbcg} shows the same information, but as a function the BCG-ICL mass. In order to be consistent with the set of observed data, we derived the amount of BCG-ICL mass 
within a given radius $R_{out}$ that is a function of halo mass (see \citealt{kravtsov18} for further details). Once again, our model is in very good agreement with the observed data. 

\begin{figure}[hbt!]
\includegraphics[scale=0.75]{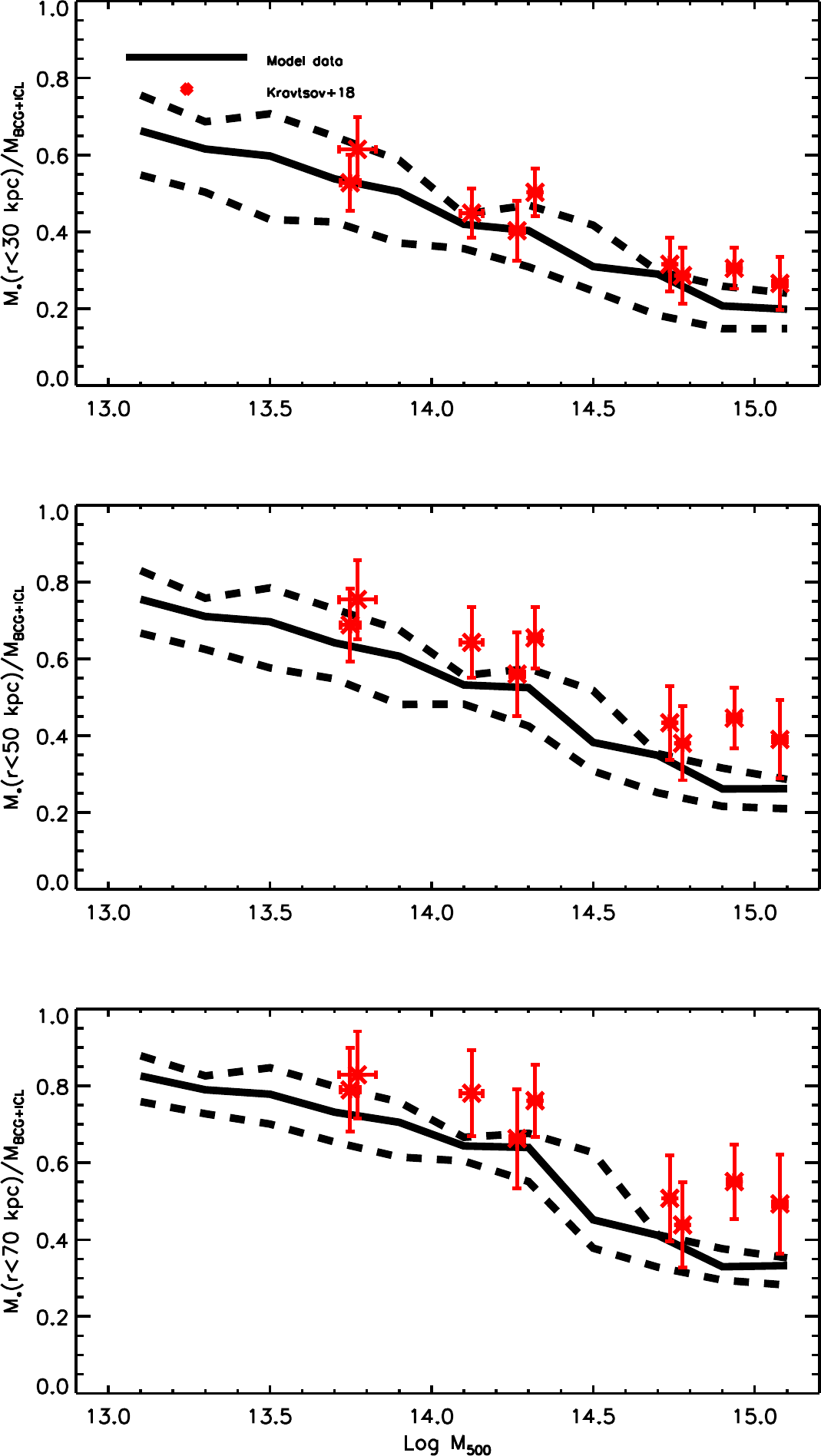}
\centering
\caption{BCG-ICL stellar mass within different apertures (30, 50 and 70 kpc) normalized to the amount of BCG-ICL mass within a given radius $R_{out}$ (see text for further details), as a function of the halo mass $M_{500}$, in the 
local universe. Dashed lines represent the 16-th and 84-th percentiles, while solid lines represent the median of the distributions. As in Fig \ref{fig:mXX__halo}, our model predictions (black lines) are compared 
with observed data (red stars) by \cite{kravtsov18} and show a good agreement. Similarly to Fig \ref{fig:ratio_halo}, in less massive haloes the BCG-ICL mass is more concentrated in the innermost regions than more 
massive haloes. The trend here is clearer because the normalization accounts for a higher amount of BCG-ICL stellar mass.}
\label{fig:mXXratio_rout_halo}
\end{figure}

In Figure \ref{fig:mXXratio_rout_halo} we plot the ratio between the BCG-ICL within the same apertures and the total BCG-ICL mass within $R_{out}$ as a function of halo mass (a similar plot can be obtained as 
a function of the BCG-ICL mass with similar information). Our model predictions (black lines) compare well with the observed data (red symbols) for a wider range of halo mass with respect to the similar plots 
in Figure \ref{fig:ratio_halo}. Overall, the model is able to capture the trend shown by the observed data, and as already seen in Figure \ref{fig:ratio_halo}, with less massive haloes having a larger concentration of 
mass in the innermost regions than more massive haloes. In this case, being the normalization higher, the trend is even more visible. However, contrary to Figure \ref{fig:ratio_halo}, here the model seems to 
be somewhat biased low rather than high. We will come back on this in Section \ref{sec:discussion}.

\subsection{The Internal BCG-ICL Mass Distribution}
\label{sec:mass_distr}

As shown in the previous subsection, our model is able to predict several observed scaling relations between the BCG-ICL mass within different apertures and halo/BCG-ICL total mass, in particular at $z=0$.
We now make use of our model to probe the mass distributions as a function of distance from the cluster centre of the three main components (bulge, disk and ICL) separately, that is the main purpose of this study. 

\begin{figure*}
\centering
\begin{subfigure}[t]{\textwidth}
\includegraphics[width=\textwidth]{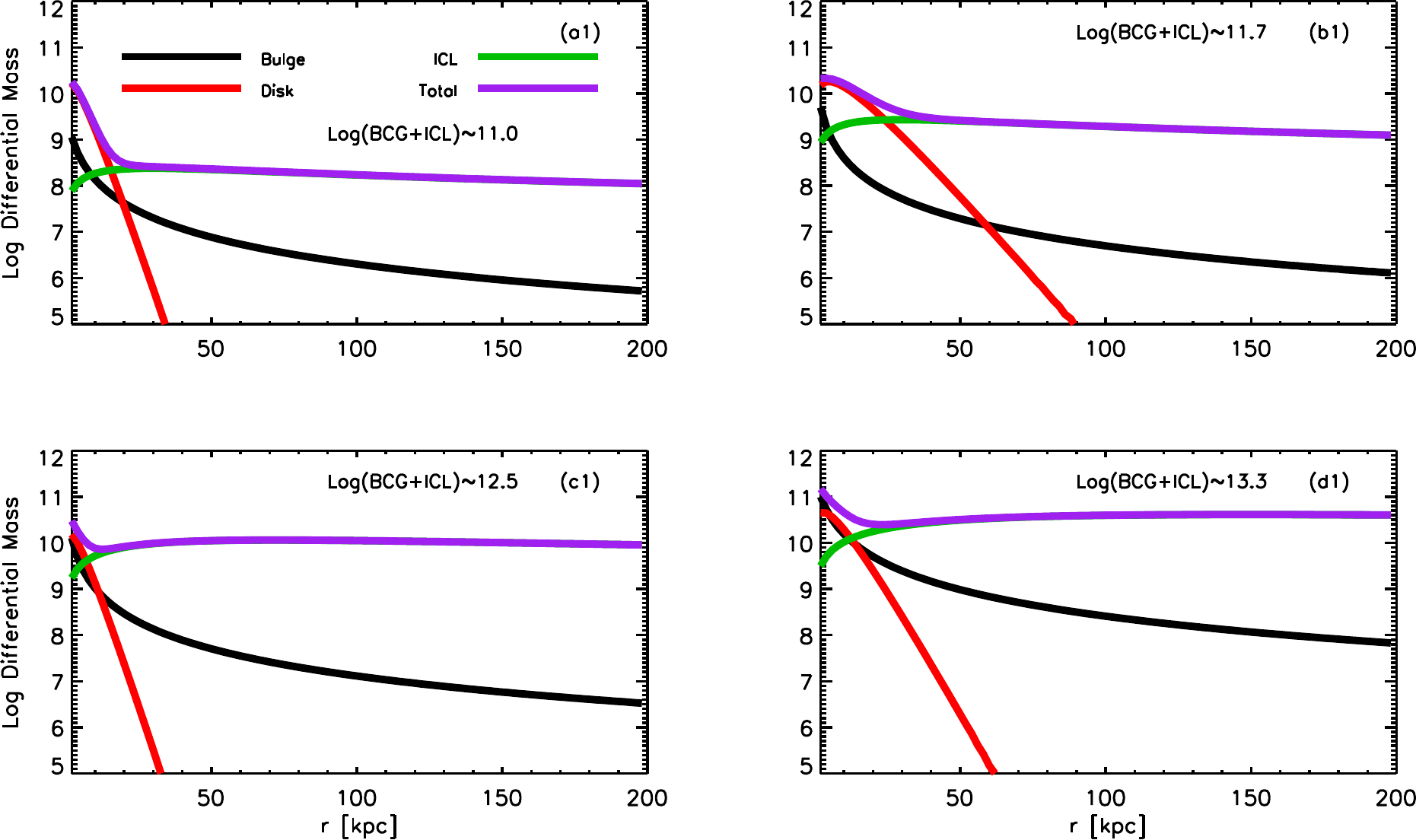} 
\end{subfigure}
 \caption{Differential stellar mass of bulge (black line), disk (red line), ICL (green line) and their sum (purple line) as a function of clustercentric distance for four BCG-ICL systems with different mass,
 as indicated in the legend of each panel. } 
 \label{fig:dm_bcgicl}
\end{figure*}
\begin{figure*}\ContinuedFloat
\begin{subfigure}[t]{\textwidth}
\includegraphics[width=\textwidth]{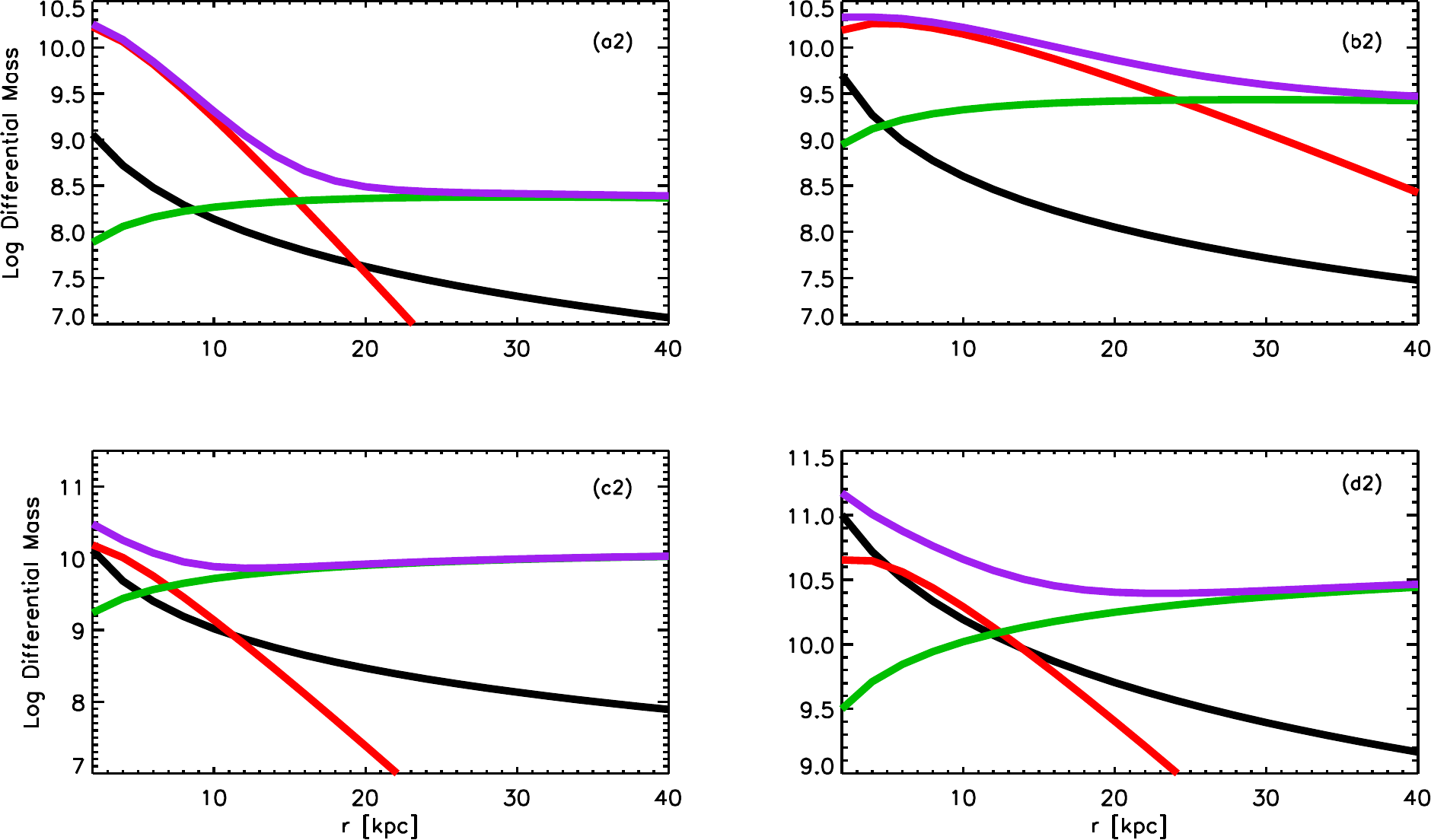} 
\end{subfigure}  
 \caption{Continued: zoom-in of the innermost kpc of each panel. Regardless of the BCG-ICL mass, our model suggests that the ICL can start to dominate the total mass in the very innermost regions, even 
 within 15-20 kpc.} 
\end{figure*}

\begin{figure*}
\centering
\begin{subfigure}[t]{\textwidth}
\includegraphics[width=\textwidth]{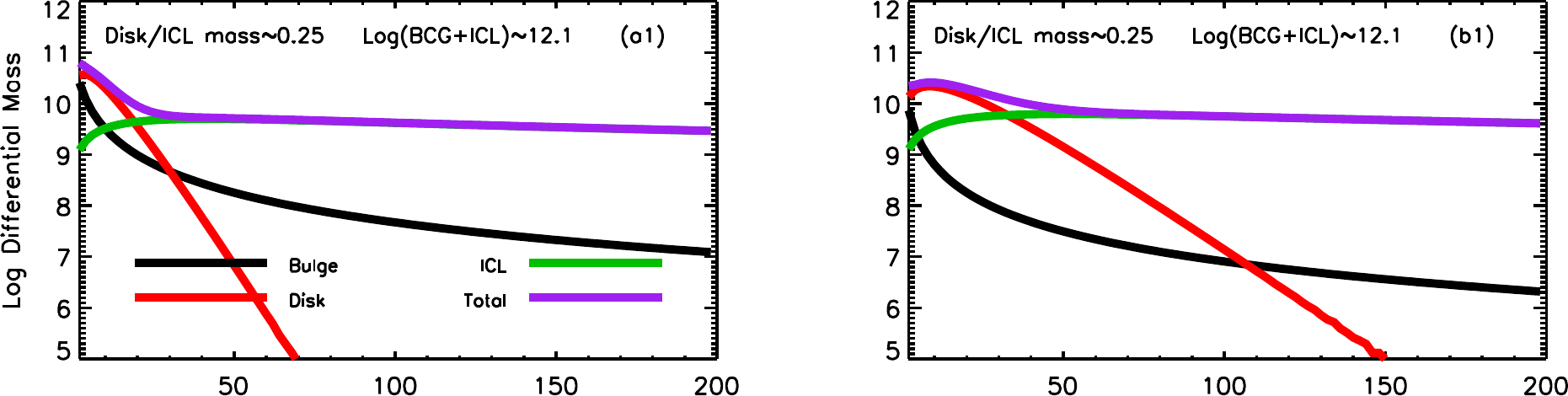} 
\end{subfigure}
\begin{subfigure}[t]{\textwidth}
\includegraphics[width=\textwidth]{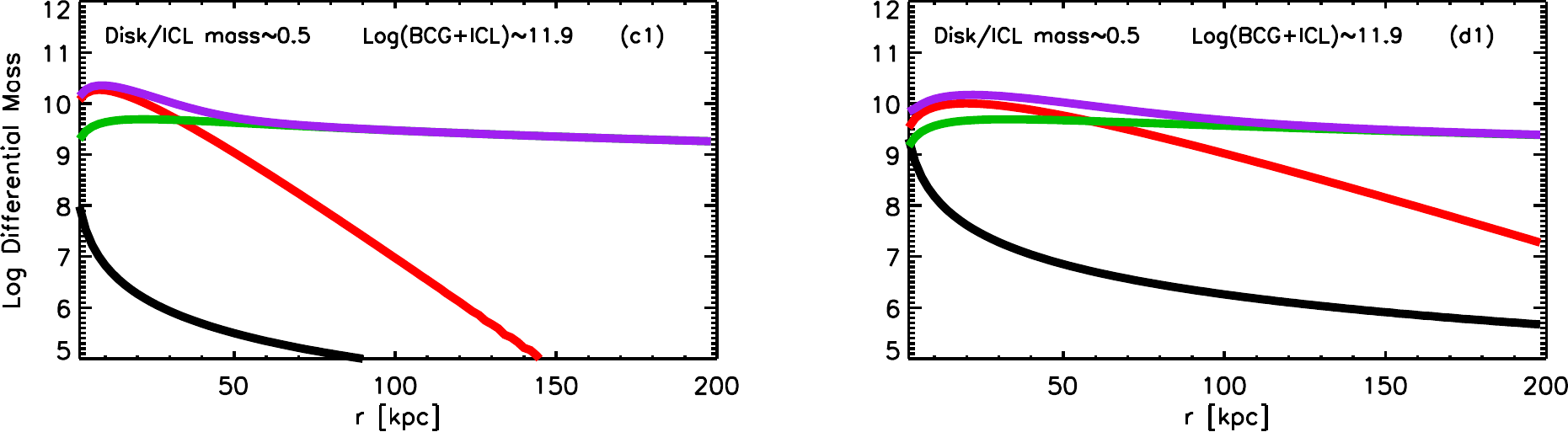} 
\end{subfigure}
 \caption{Same as Fig \ref{fig:dm_bcgicl} for galaxies with similar BCG-ICL stellar mass, but different disk/ICL mass ratio (0.25 in the upper panels and 0.5 in the bottom panels).} 
 \label{fig:di_ratio}
\end{figure*}
\begin{figure*}\ContinuedFloat
\begin{subfigure}[t]{\textwidth}
\includegraphics[width=\textwidth]{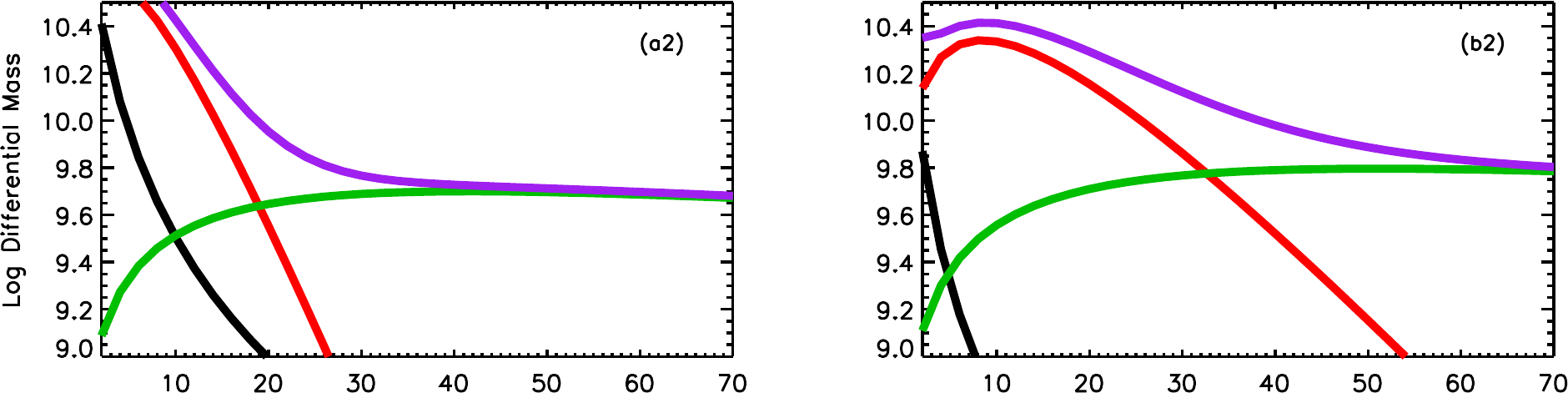} 
\end{subfigure}
\begin{subfigure}[t]{\textwidth}
\includegraphics[width=\textwidth]{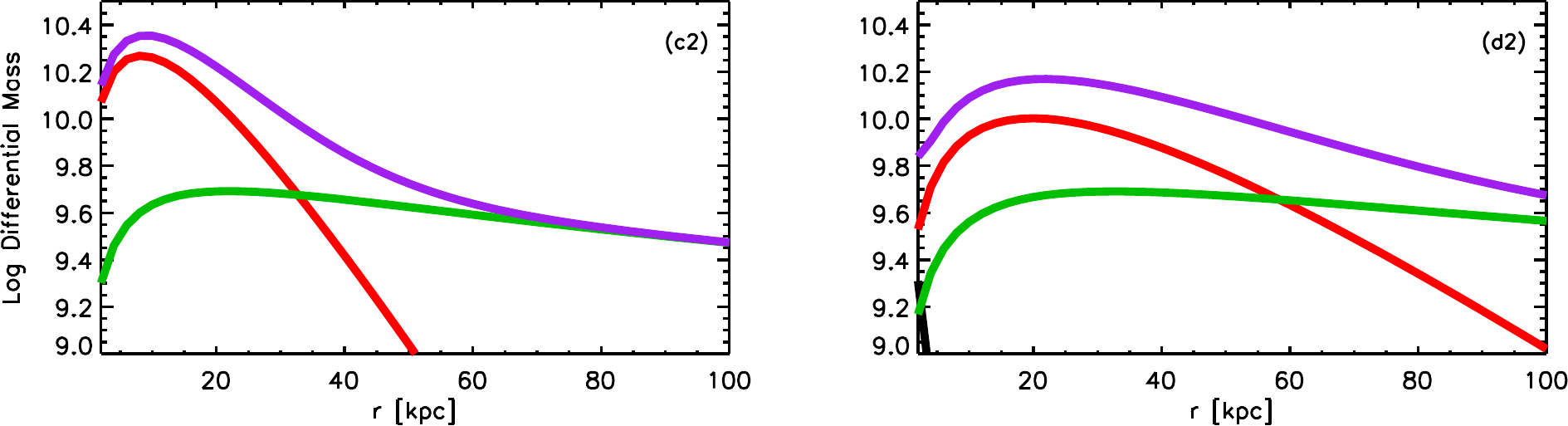} 
\end{subfigure}  
 \caption{Continued: zoom-in of the innermost kpc of each panel. When BCGs have a more prominent disk compared to the amount of ICL associated to them (bottom panels), the ICL starts to dominate the total mass 
 at farther distances ($>70$ kpc for c2 and $>100$ for d2).} 
\end{figure*}

\begin{figure*}
\centering
\begin{subfigure}[t]{\textwidth}
\includegraphics[width=\textwidth]{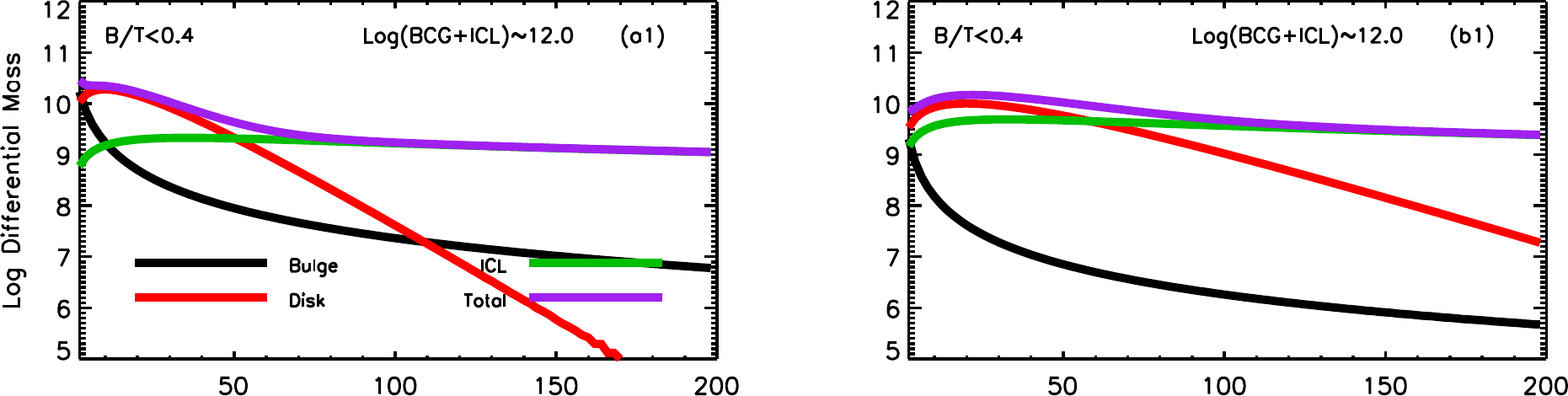} 
\end{subfigure}
\begin{subfigure}[t]{\textwidth}
\includegraphics[width=\textwidth]{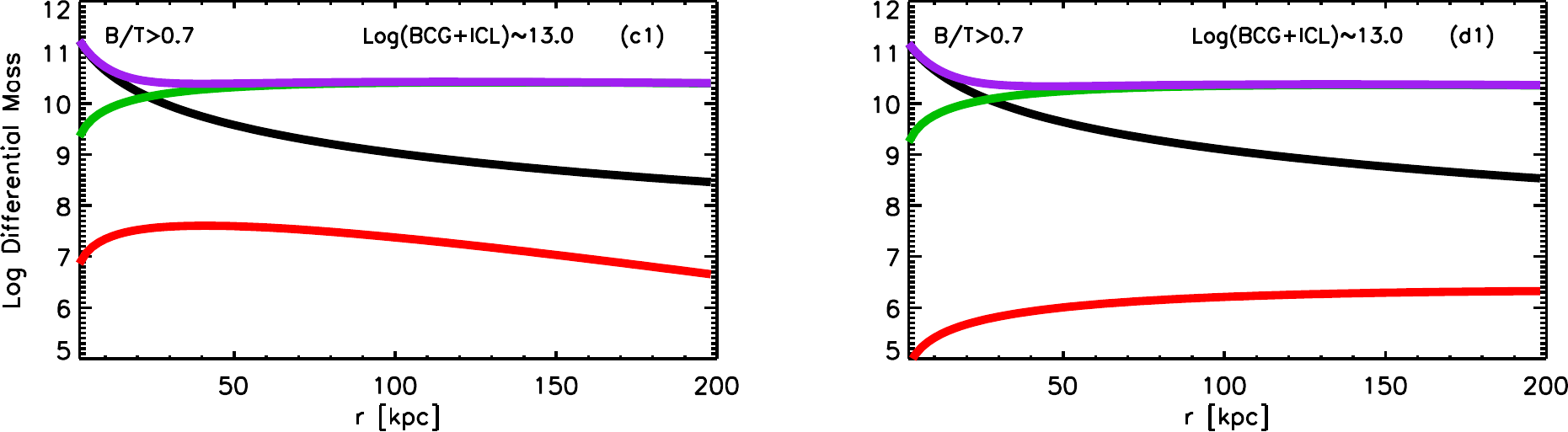} 
\end{subfigure}
 \caption{Same as Fig \ref{fig:di_ratio} for galaxies with similar BCG-ICL stellar mass, but different bulge/total mass ratio (interpreted as bulge/(bulge+disk)) as indicated in the legend of each panel.} 
 \label{fig:bt_ratio}
\end{figure*}
\begin{figure*}\ContinuedFloat
\begin{subfigure}[t]{\textwidth}
\includegraphics[width=\textwidth]{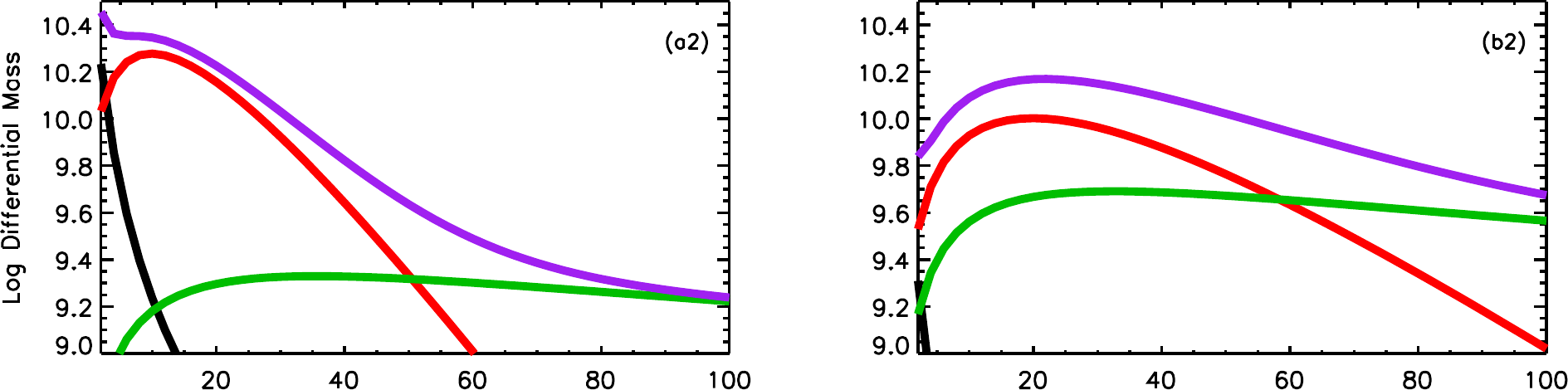} 
\end{subfigure}
\begin{subfigure}[t]{\textwidth}
\includegraphics[width=\textwidth]{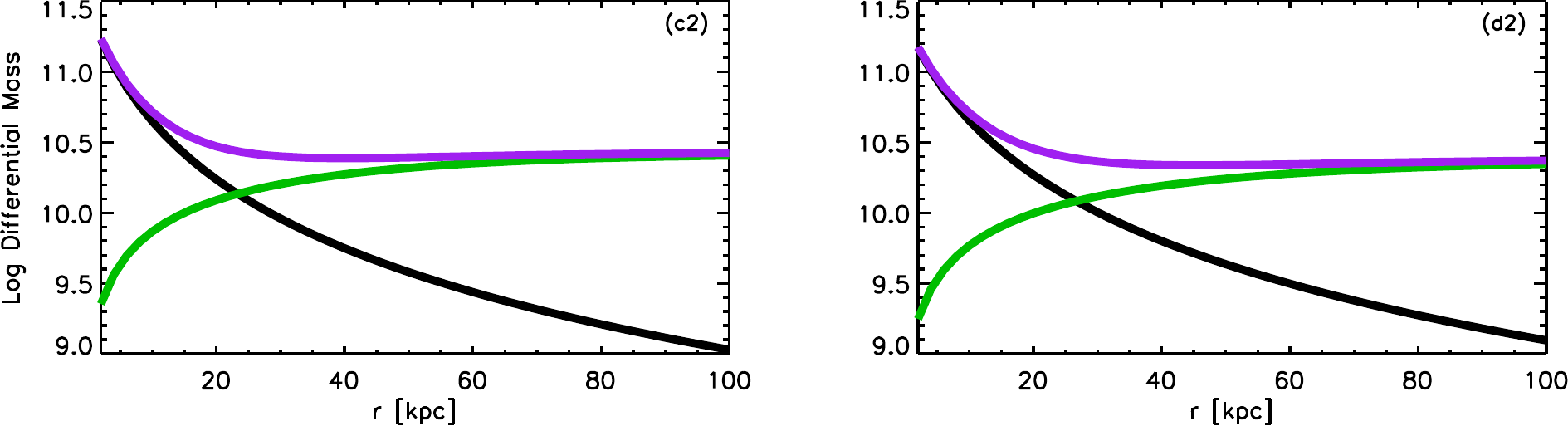} 
\end{subfigure}  
 \caption{Continued: zoom-in of the innermost kpc of each panel. When BCGs are more ``disky" (upper panels), the ICL starts to dominate the total mass at farther distances ($>60$ kpc).} 
\end{figure*}

The first four panels of Figure \ref{fig:dm_bcgicl} show the diffferential stellar mass of bulge (black line), disk (red line), ICL (green line) and their sum (purple line) as a function of clustercentric 
distance for four BCG-ICL systems with different mass, ranging from $\log (BCG+ICL) \sim 11.0$ (panel a1), to $\log (BCG+ICL) \sim 13.3$. These case studies have been chosen almost randomly from our sample, with the 
only condition of being in the lowest quartile of the mass range (a1), some percentiles before and after the median (b1 and c1), and in the highest quartile (d1). No other condition regarding the mass in the bulge 
or disk has been chosen, neither about the morpholgy of the BCGs, or halo mass. For these ``random" case studies, it is quite evident that the total BCG+ICL stellar mass does not give a real indication of which 
component (and from where) starts to dominate the total distribution. This is more clear in the last four panels (a2, b2, c2 and d2) of the same figure, which are a zoom-in of the previous four of the innermost 
regions. In all cases the bulge, as expected, drops very fast, while disk and ICL dominate the total distributions at different clustercentric distances, with the only exception given by d2 for which the mass in the 
disk is small compared to that in the bulge. Very interestingly, the ICL starts to dominate the total distribution at very small radii, even at 15-20 kpc. 

We now move the attention to other four case studies: two pairs of BCG-ICL systems with similar mass and disk/ICL mass ratio. Their differential mass for each component as a function of distance is shown in Figure 
\ref{fig:di_ratio}, four a pair with $\log (BCG+ICL) \sim 12.1$ and disk/ICL mass $\sim 0.25$ (panels a1 and b1), and a pair with $\log (BCG+ICL) \sim 11.9$ and disk/ICL mass $\sim 0.5$ (panels c1 and d1). Again, 
the four BCGs have been chosen in a random manner among those in our sample, with the only conditions of having same total mass and disk/ICL mass ratio for members of each pair, and similar total mass but different 
disk/ICL mass ratio for pair-pair members. For these case studies, the picture looks quite different with respect to the previous ones. Indeed, all case studies have a discreet disk/ICL mass ratio, that translates in 
a more relevant importance of the disk compared to the ICL. From panel a1 to d1 (similar mass but increasing disk/ICL mass ratio) we can see that the disk is increasingly important, and this makes the ICL dominate the 
total distribution at larger radii than seen before. This trend is more visible in the zoom-in shown in the last four panels of the same figure. As expected, the bulge drops even faster because these are BCGs with a
prominent disk, and the importance of the disk with respect to the ICL makes all the difference. In the two cases where the disk is a quarter of the ICL (a2 and b2), the latter starts to dominate at radii $\sim 50-60$
kpc, but when the disk is half of the ICL, the distribution of the latter meets the total one at farther distances, $>70$ kpc in the case c2 and $>100$ kpc in case d2. 

It appears clear that in disk-like BCGs the ICL is less likely to dominate from the innermost regions, while for bulge-like BCGs, given the fact that the bulge drops faster than the disk, it is more likely. We 
address this point in Figure \ref{fig:bt_ratio}, where we show other two pairs of case studies with similar mass and bulge/BCG mass ratio and plot again the differential mass of each component as a function of 
distance. The first two cases (a1 and b1) are two disk-like BCGs with B/T$<0.4$ and $\log (BCG+ICL) \sim 12.0$, and the second two cases (c1 and d1) are two bulge-like BCGs with B/T$>0.7$ and $\log (BCG+ICL) \sim 13.0$.
For these four cases the conditions were medium size disky BCGs and two bulge-like BCGs on the last quartile of the mass range. The reasons behind our choices is that we want to see the difference bewteen BCGs of 
different morphology and, at the same time, whether or not in a bulge-like BCG the ICL can dominate the total distribution in a similar way as in a smaller disky BCG. In the first two cases the disk is the component 
that ''competes" with the ICL and the bulge does not contribute much, while in the last two cases the bulge makes the difference and an almost absent disk does not contribute. The main features of these plots can be 
seen in the zoom-in provided in the last four panels of the same figure. Similar to previous cases where the disk is important with respect to the ICL, the latter component starts to dominate the total distribution at 
large distances (over 100 kpc), but the interesting aspect is that in massive bulge-like BCGs the ICL can be important only at distances beyond 60-70 kpc.

With all the previous case studies of different BCG-ICL systems we have shown that the ICL can be the most important component, and so not neglected, even at very small clustercentric distances such as 10-20 kpc. 
Moreover, the morpholgy of the BCGs, the importance of the mass in the disk with respect to that of the ICL as well as the relative importance of the total BCG mass with respect to that of the bulge, are parameters 
that weigh on the distance at which the ICL begins to be the only component at play, and the range of distances can be as wide as $\sim 100$ kpc. In the next section we will further discuss all our results, but mainly 
those of the second part of our analysis in light of the common assumptions that are generally made when distinguishing between BCG and ICL.

\section{Discussion}
\label{sec:discussion} 

One of the most serious problems in studying the ICL come directly from its own definition (e.g., \citealt{rudick11,cui14,jimenez16,montes18,zhang19}). Indeed, different authors use different definitions and this happens in 
both theoretical and observational sides, which makes comparisons among different studies quite complicated. Measuring the ICL is already a hard task, and attempts to separate it from the BCG complicate the job even more, 
simply because the two components are overlapped in the region dominated by the BCG and the transition between them is supposed to be smooth. Theoretical speaking, it is quite easy to define the ICL because in both 
semi-analytics and hydrosimulations it is possible to take advantage of information not allowed to observers (see, for example, \citealt{rudick11} for a detailed review of the most common methods). 

On the other hand, in observations, the picture is rather different. Among the years, mainly two are the methods that have been constantly used in observational studies. The easiest way is to apply some cut 
in surface brightness after removing all the contribution from other sources, such as satellite galaxies and sky. All the light fainter than the cut is defined as ICL, and the rest is given by the BCG 
(e.g \citealt{burke15,montes18,spavone20,iodice20} and references therein). As mentioned above, clearly this method might severely underestimate the contribution of the ICL in the innermost regions where it overlaps with the BCG, and thus, might 
overestimate that of the BCG. This approach is similar, methodologically speaking, to assume a given distance from the center where the light profile drops and follows a plateau, and assume that from there on the ICL 
is the only component. The other common approach is to use functional forms as we have done in CG20 (and in this paper), that allows to distribute the light of BCGs and ICL (e.g., \citealt{kravtsov18,zhang19,kluge20,montes21}).
To quote a few them, \cite{montes21} use GALFIT (\citealt{peng02}) to fit a double (one for the BCG and one for the ICL) Sersic (\citealt{sersic68}) profile. Similarly to \cite{kravtsov18}, \cite{zhang19} used a 
combination of three Sersic profiles: one for the inner core, one for the bulge and one for the ICL. Either with a double or triple Sersic profile, the different components dominate a different distance from the centre. 
We will come back on this point below.

\subsection{Is Our Model Reliable?}
\label{sec:mod_rel}

In Section \ref{sec:model_set} we presented the analysis done in order to set our model, by comparing our predictions with observational data by \cite{demaio20} up to redshift $z\sim 1.5$, and by \cite{kravtsov18} at the
present time. Overall we find a compelling agreement with observations, but a key point it is worth discussing deeper. When we plot the $M_{10-100}-M_{500}$ relation at different redshifts, we find a gap of about $\sim 0.3$ 
dex between our predictions and the observed ones at redshift $z\gtrsim 0.5$. A similar gap has been found in CG20 when we included the innermost 10 kpc. The question that rises naturally is: Is this gap real or explainable
by invoking intrisic differences between the model and the observation? For real we mean that, as a matter of fact, it implies an evolution in mass since redshift $z\sim 0.5$ of a factor $\sim 2$ that it is not seen in 
the observed data. In CG20 we have addressed this point. As a possible cause for the gap we have argued that, while our predictions are exactly at that particular redshifts, $z\sim 0.5$ and present time, the observational
data by DeMaio et al. span a wider range of redshift with median at $z\sim 0.5$. Another possible reason is that our BCG+ICL mass are actually lower than observed at that redshift, which could be due to the difficulty in 
dectecting the stellar mass at higher redshift given by all the masking done for separating the source from the sky. It might also be that our model, at least in part, underestimates the BCG+ICL mass at 
$z \gtrsim 0.5$. 

According to DeMaio et al., it is possible to reconcile the theoretical prediction of our model (since it predicts a rapid evolution of the ICL at late times, \citealt{contini14}) with their data if most of the ICL 
growth actually occurs at $r>100$ kpc, which is a very reasonable explanation. However, here we try to investigate the innermost tens kpc by assuming a mass profile for every component of a BCG+ICL system, and a direct 
comparison with their data shows that, although minor, our model does show some evolution within 100 kpc from $z\sim 0.5$ to the present time. The current state of affairs does not allow us to answer to the question 
above, because the gap is small for making final statements. Nevertheless, while we suggest the need to have more data at $z\sim 0.5$ in order to get a more complete picture and a fairer comparison, we surely stress that 
the profiles we assumed might not completely describe the internal BCG+ICL mass distribution at any redshift, even though they do at the present time. 

DeMaio et al. correctly explain that the BCG+ICL gains stellar mass in the range 10-100 kpc more rapidly than within 10 kpc by invoking the ``inside-out'' scenario (\citealt{vandokkum10,patel13,bai14,vanderburg15,zhang16,zhang19} 
and others) for the growth of the stellar mass distribution. Despite the issue discussed above, our model agrees well with this scenario. Indeed, albeit our predictions are slightly biased low with respect to the observed 
data, the trend is reproduced, i.e., as a cluster grows in halo mass, it gains more stellar mass from the outer regions (see Figure \ref{fig:ratio_halo}). 

We then conclude that our model is reliable at least at the present time, while it needs more testing at higher redshift when more data are available. Hence, a model with a Jaffe profile for the bulge, an exponential one 
for the disk, and an NFW distribution of the ICL mass can be a good description of the whole BCG+ICL system. In the light of this, we now discuss the main point of this paper, that is, at what distance from the centre 
the ICL starts to dominate the mass distribution. 

\subsection{Where Does the ICL Dominate the Mass Distribution?}
\label{sec:icl_dom}

In Section \ref{sec:mass_distr} we have investigated the mass distribution of several case studies of BCG+ICL systems by looking at the distribution of each component. We have shown that the distance from which the 
ICL starts to dominate the total distribution strongly depends on the BCG properties, such as its morphology and whether or not it has an extended disk. The radius at which the ICL dominates the total distribution 
(for the sake of simplicity let's call it $R_{transition}$ hereafter) can be very short, around 10-20 kpc, but it might reach even over 100 kpc, depending on the particular case. The total BCG+ICL stellar mass does not 
seem to play an important role (Figure \ref{fig:dm_bcgicl}), while the relative contribution of the disk with respect to the ICL (Figure \ref{fig:di_ratio}) and that of the bulge with respect to the total BCG mass 
(Figure \ref{fig:bt_ratio}) do.

The morphology, bulge/disk-like BCG, has a major role in determining from where the ICL becomes the dominant component. It is plausible and somewhat expected that BCG+ICL systems where the BCG has an extended disk had 
larger $R_{transition}$, and bulge-like BCGs, depending on the mass of their bulge, more or less shorter $R_{transition}$. Having or not an extendend disk depends mostly on the merger history of the particular BCG, mainly 
whether or not it has experienced a recent major merger, which totally destroys the disk. However, the disk can further form again if there is enough cold gas and time. Mergers, on the other hand, contribute also on the 
production of the ICL, i.e., the more the number of mergers experienced, the more the mass that will end up in the ICL component. We can distinguish among three different case scenarios that might explain the BCG+ICL distribution for 
BCGs (a) bulge dominated (the most common), (b) spheroidals with similar bulge and disk components, and (c) disk-like BCGs (rare objects).

In the case (a), such BCGs either have had a troubled merger history ceasing the star formation before a new disk could grow again (a1), or they had a very recent major merger and no time available to grow a new disk (a2). 
In the case (a1) the amount of ICL is expected to be large, given the fact that mergers do contribute to the ICL and it is very likely that stellar stripping would play an important role when satellite galaxies are 
merging. In the case (a2), most depends whether the ICL formed through stellar stripping at late times. Case (b) represents a sort of middle ground between (a) and (c). These BCGs might have had mergers in the past (and 
so formed ICL), but no recent major merger so that the disk could grow again. For disk-like BCGs, i.e. case (c), it is very likely that they did not have any major merger in the recent past (or at all), and they evolved 
mainly through smooth accretion/minor mergers and via star formation until gas has been available. However, such BCGs are very rare and constitute a class of BCGs for which their ICL should not be expected to be the 
relevant component.  

Given these three possible scenarios for the formation and evolution of BCG+ICL systems, $R_{transition}$ spans a wide range. It is the shortest for systems where the BCG is a bulge-like galaxy in the low end of 
the mass rank, and the highest where the BCG has a prominent disk. Overall, considering our case studies, we find that 15 kpc $\lesssim R_{transition}\lesssim 100$ kpc, but can be even larger than 100 kpc in very 
particular cases. Let's discuss our results in the context of the most recent observational evidence.

Recently, \cite{montes21} focused on a deep study of the ICL of the Abell 85 galaxy cluster by using archival data taken with Hyper Suprime-Cam on the Subaru Telescope. They measured the BCG+ICL surface brightness 
profile out to around 215 kpc in the g and i bands, and noticed that, at $\sim 75$ kpc, both the surface brightness and color profiles become shallower. The interpretation of this result is that the ICL starts to dominate 
at that radius. Not only, in agreement with the prediction of our model (\citealt{contini14,contini19}) the color profile at radius beyond 100 kpc is consistent with that of massive satellite galaxies, suggesting that 
the stripping of these objects is responsible for the build-up of the ICL. Their $R_{transition} \sim 70$ kpc is also in good agreement with $R_{transition} \sim 80$ kpc by \cite{zibetti05}, who found a similar 
flattening of the profile at that radius. It must be noted that the value of $R_{transition}$ by Zibetti et al. is the result of a stacking of about 700 galaxy clusters taken by the SDSS. This value of the transition 
radius at which the ICL becomes the dominant component agrees fairly well with works (e.g., \citealt{gonzalez05,seigar07,iodice16,gonzalez21}) that showed that BCG+ICL systems have such a break in the profile in the range 60-80 kpc.

Similar numbers have been found by \cite{zhang19} from the analysis of about 300 galaxy clusters at redshift $0.2<z<0.3$. These authors described the BCG+ICL radial profile by using a triple Sersic model and found a 
dominant core in the innermost 10 kpc, a bulge between 30 and 100 kpc and the ICL that dominates the distribution beyond 200 kpc, but already outside 100 kpc the profile transitions from the BCG to the ICL. They 
suggested that this transition is a potential cut between the two components, in good agreement with \citealt{pillepich18} who noticed that simulated haloes more massive than $5\cdot 10^{14} \, M_{\odot}$ have more ICL 
mass outside 100 kpc than within.

All the values of the transition radius, i.e. the radius at which the ICL starts to become the dominant component, reported above are comparable with the range predicted by our model. We stress again that, a wide range 
of possible values of $R_{transition}$ is preferable to a narrow one for the reasons discussed above. It strongly depends on the morphology of the BCG, of the amount of ICL with respect to the other components and, not 
as last, to the dymanical state of clusters, which reflects the merging history of the BCG+ICL system and so its evolution with time.

\section{Conclusions}
\label{sec:conclusions}

Following the analysis done in our recent paper (\citealt{contini20}), we have further improved our model by assuming that the radial mass distribution of a BCG+ICL system can be described by the sum of three profiles:
a Jaffe profile for the bulge, an exponential disk and a modified NFW for the ICL. We have taken advantage of a wide sample of simulated galaxy groups and clusters from our state-of-art semi-analytic model to study 
the innermost $\sim 100$ kpc of BCG+ICL systems. Unlike the model used in our previous study, the new one offers the opportunity to go below 100 kpc since bulge and disk are treated separately. The main goals of 
this study were to confirm that our model can reproduce several observed relations between the stellar mass at a given radius and the total BCG+ICL/halo mass, at different redshift and, give an idea of where the 
transition between the BCG and the ICL happens, i.e. when the ICL component starts to dominate the total distribution. In the light of the analysis done and the overall discussion, we conclude the following:
\begin{itemize}
 \item the radial mass distribution of a BCG+ICL system can be described by the sum of three profile: a Jaffe profile for the bulge, an exponential disk and a NFW-like profile for the ICL. The model can reproduce 
       several observed scaling relation between the BCG+ICL stellar mass within a given aperture and the BCG+ICL/halo mass, at different redshift with an overall precision below 0.05 dex at $z=0$ and 0.3 dex at 
       $z\gtrsim 0.5$ along the wide stellar mass/halo mass ranges probed. 
 \item In good agreement with recent observational studies, our model predicts that $R_{transition}$, i.e. the radius at which the radial distribution of BCG+ICL systems, transitions from the BCG to the ICL, ranges 
       from $\sim 15$ kpc to $\sim 100$ kpc. The transition radius, however, depends on BCG properties such as its morphology, bulge/disk-like galaxy, on the amount of ICL with respect to the other components, and 
       to the dymanical state of the group/cluster.
\end{itemize}

In a future study we aim to better set the model at higher redshifts by investigating the parameter $\gamma$, that links the concentration of the ICL to that of the halo, and its relation between halo mass and redshift.
In order to do that, we need more observational data at $z\gtrsim 0.5$ where our model seems to show some bias with respect to the data used in this study.

\section*{Acknowledgements}
The authors thank the anonymous referee for his/her constructive comments which helped to improve the manuscript.
This work is supported by the National Key Research and Development Program of China (No. 2017YFA0402703), 
and by the National Natural Science Foundation of China (Key Project No. 11733002). 
\label{lastpage}

\end{document}